\documentclass[preprint,showpacs,preprintnumbers,amsmath,amssymb]{revtex4}
\usepackage{epsfig}
\usepackage{graphicx}
\usepackage{ulem}
\usepackage{color}
\definecolor{My_red}{cmyk}{0.00,1.00,1.00,0.20}

\def\be{\begin{equation}}
\def\ee{\end{equation}}
\def\bea{\begin{eqnarray}}
\def\eea{\end{eqnarray}}
\def\nn{\nonumber}
\def\spur{\not \!}
\begin{document}
\title{Can contribution from  magnetic-penguin operator with real photon to
$B_s\to \ell^+\ell^-\gamma$ in the standard model  be neglected? }
\author{Wenyu Wang, Zhao-Hua Xiong and Si-Hong Zhou}
\affiliation{Institute of Theoretical Physics, College of Applied Science,
     Beijing University of Technology, Beijing 100124, China}
\date{\today}
\begin{abstract}
Using  the $B_s$ meson wave function extracted from non-leptonic $B_s$ decays,
we reevaluate the rare decays $B_s\to \ell^+\ell^-~\gamma,~(\ell=e,\mu)$
in the standard model, including two kinds of contributions from
magnetic-penguin  operator with virtual and real photon.
We find that the contributions from magnetic-penguin
operator $b\to s\gamma$ with real photon to the exclusive decays,
which is regarded as to be negligible in previous literatures,
are large, and the branchings of  $B_s\to \ell^+\ell^-\gamma$ are nearly enhanced
by a factor 2. With the predicted branching ratios at order of $10^{-8}$,
it is expected that the radiative dileptonic decays   will be detected
in the LHC-b and B factories in near future.
\end{abstract}
\maketitle

\section{Introduction}
The standard model (SM) of electroweak interaction has been
remarkably successful in describing physics below the Fermi
scale and is in good agreement with the most experiment data. Thanks
to the efforts of the B factories and LHC, the exploration of
quark-flavor mixing is now entering a new interesting era.
Measurements of rare B mesons decays such as  $B \to X_s \gamma$,
$B\to X_s\ell^+\ell^-(\ell=e,\mu)$ and  $B_s\to \mu^+\mu^-$,
$B_s\to \ell^+\ell^-\gamma$  are likely to provide sensitive
test of the SM. In fact, these decays, induced by the flavor
changing neutral currents  which occur in the SM only at loop level,
play an important role in testing higher order effects of SM
and in searching for the physics beyond the SM \cite{Aliev97,Xiong01}.
Nevertheless, these processes are also important in determining
the  parameters of the SM and some hadronic parameters in QCD,
such as the CKM matrix elements, the meson decay constant $f_{B_s}$,
 providing information on heavy meson wave functions~\cite{li}.

The rare B inclusive radiative decays  $B \to X_s \gamma$ and
$B\to X_s\ell^+\ell^-(\ell=e,\mu)$ as well as the exclusive decays
$B_s\to \mu^+\mu^-$  have been studied extremely at the leading
logarithm order~\cite{BLOSM} and high order in the SM~\cite{BHOSM}
and various new physics models. In previous works, prediction for
the exclusive decays $B_s\to\ell^+\ell^-\gamma$ have been carried out
by using the light cone sum rule~\cite{Aliev97,Xiong01}, the simple
constituent quark model~\cite{Eilam97},  and the B meson distribution
amplitude extracted from non-leptonic B decays~\cite{LU06}.
long distance QCD effects describing
the neutral vector-meson resonances $\phi$ and $J/\Psi$ family
have received special attention in  \cite{Melikhov04,Kruger03,Nikitin11}.
At parton level, $B_s\to\ell^+\ell^-\gamma$ decays
have been thought to be obtained from decay
$b\to s\ell^+\ell^-\gamma$, and further, from $b\to s\ell^+\ell^-$ directly.
To achieve this, a necessary work is attaching real photon to any charged
internal and external lines in the Feynman diagrams of $b\to s\ell^+\ell^-$
with two statements: i) Contributions  from the attachment of photon to any
charged internal propagator are regraded as to be strongly suppressed and can be neglected
safely~ \cite{Aliev97,Xiong01,LU06,Eilam97};
ii) Contributions from the attachment of real photon with magnetic-penguin
vertex to any charged  external lines are always neglected~\cite{Aliev97,Xiong01}
or stated  to be negligibly small~\cite{LU06}.  Here we would like to address
that the conclusion of the first statement is correct, but the explanation
is not as what it is described~\cite{Dong}.  The second statement seems to be questionable,
for that the pole of propagator of the charged line attached by photon may enhance
the decay rate greatly which make some diagrams can not be neglected in the
calculation. Since the weak radiative B-meson decay is well known to be a sensitive probe of new physics,
it is essential to calculate the Standard Model value of its branching ratio
as precisely as possible.  Although the second contribution has been calculated in
Ref. \cite{Melikhov04}, it mainly concentrated on the long distance effects of the
meson resonances, whereas the short distance contribution  which was incompletely analyzed.

In this letter, we will concentrate on the short
distance contribution to $B_s\to \ell^+\ell^-\gamma$ and check whether the contribution from
magnetic-penguin operator with {\it real} photon to $B_s\to \ell^+\ell^-$ is negligible
or not, and give some remarks including a comparison with other works.
The paper is organized as follows. In sec. \ref{sec:hami}, we present the
detailed calculation of exclusive decays $B_s\to\ell^+\ell^-\gamma$, including
full contribution from magnetic-penguin operator with real photon.
Sec. \ref{sec:num} contains the numerical results and comparison with previous works, and
the conclusions are  given in sec. \ref{sec:con}.

\section{The calculation}\label{sec:hami}
In order to simplify the  decay amplitude for $B_s\to \ell^+\ell^-\gamma$,
we have to utilize the $B_s$ meson wave function, which is not known from
the first principal. Fortunately, many studies on non-leptonic $B$
\cite{bdecay,cdepjc24121} and $B_s$ decays \cite{bs} have constrained
the wave function strictly.  It was found that the wave function has form
 \begin{equation}
\Phi_{B_s}= (\not \! p_{B_s} +m_{B_s}) \gamma_5 ~\phi_{B_s}
({x}), \label{bmeson}
\end{equation}
where the distribution amplitude $\phi_{B_s}(x)$ can be expressed as~\cite{form}:
\begin{equation}
\phi_{B_s}(x) = N_{B_s} x^2(1-x)^2 \exp \left( -\frac{m_{B}^2\ x^2}{2
\omega_{b_s}^2}  \right)
\label{phib}
\end{equation}
with $x$ being  the momentum fractions shared
by $s$ quark in $B_s$ meson.
 The  normalization constant $N_{B_s}$ can be determined by comparing
\begin{eqnarray}
\langle 0\left|\bar s \gamma^{\mu}\gamma_{5}b\right|B_s\rangle=i\int_{0}^{1}\phi_{B_s}(x)dx
{\rm Tr}\left[\gamma^{\mu}\gamma_{5}(\spur p_{B_s}+m_{B_s})\gamma_{5}\right]dx
=-4ip_{B_s}^{\mu}\int_{0}^{1}\phi_{B_s}(x)dx
\end{eqnarray}
with
\begin{eqnarray}
\langle0\left|\bar s\gamma^{\mu}\gamma_{5}b\right|B_s\rangle=-if_{B_s}p_{B_s}^{\mu},
\end{eqnarray}
the  $B$ meson decay constant $f_{B_s}$ is thus determined by
the condition
\begin{eqnarray}
\int_{0}^{1}\phi_{B_s}(x)dx=\frac{1}{4}f_{B_s}.
\end{eqnarray}

Let us start with the quark level processes $b\to s\ell^+\ell^-$.
They are subject to the QCD corrected effective weak Hamiltonian,
obtained by integrating out heavy particles, i.e., top quark, higgs, and $W^\pm,\ Z$ bosons:
\begin{eqnarray}
{\cal H}_{eff}(b\to s\ell^+\ell^-)&=&-\frac{\alpha_{em}{G_F}}{\sqrt{2}\pi}V_{tb}V^*_{ts}
\left\{\left[-\frac{2{C^{eff}_7}{m_b}}
{q^2}\bar{s}i\sigma^{\mu\nu}q_\nu P_Rb
+C^{eff}_9\bar{s}\gamma^\mu{P_L}b\right]\bar{\ell}\gamma_\mu{\ell}\right.\nn\\
&&\left.+C_{10}(\bar{s}\gamma^\mu{P_L}b)~
\bar{\ell}\gamma_\mu\gamma_5{\ell}\right\},
\label{hami}
\end{eqnarray}
where $P_{L,R}=(1\mp\gamma_5)/{2}$,\ $q^2$
is  the dilepton invariant mass squared. The QCD corrected Wilson
coefficients $C_7^{eff}$, $C^{eff}_9$
and $C_{10}$  at $\mu=m_b$ scale can be found in Ref.~\cite{Misiak93}.

\begin{figure}[htbp]
\begin{center}
\scalebox{1.2}{\epsfig{file=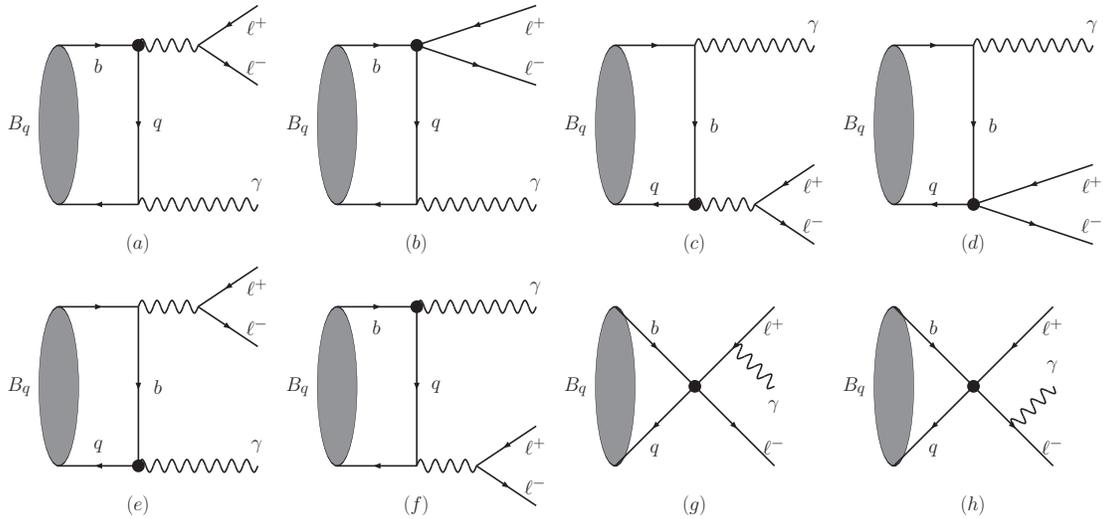}}
\caption{Feynman diagrams for $B_q\to \ell^+\ell^-\gamma~(q=d,s)$
in the SM.  The black dot in (a), (c) and (e), (f)
stands for  the magnetic-penguin operator $O_7 $ with virtual
and real photon, respectively, and the black dot in (b),(d),(g)
and (f) denotes  operators $O_9$ and $O_{10}$.}
\label{fig1}
\end{center}
\end{figure}

If an additional photon line is attached to any of the charged lines
in diagrams contributing to the  Hamiltonian above, we will have the radiative leptonic decays
$b\to s\ell^+ \ell^-\gamma$.  Therefore, there
are two kinds of diagrams: photon connecting to the internal propagators,
and photon connecting to the external line. As addressed in the introduction,
the contribution from the first kind of diagrams is neglected safely.
Now we will only consider the second category of diagrams which are displayed in Fig.~\ref{fig1}.
At first, we recalculate the diagrams (a)-(d) in Fig.~\ref{fig1} with
photon emitted from the external quark lines $b$ or $s$ by using
the $B_s$ meson wave function extracted from non-leptonic $B_s$ decays.
Note that these diagrams are already studied in previous literatures and  considered
as giving the dominant contribution to $B_s\to \ell^+\ell^-\gamma$.
At parton level, the amplitudes for  transition $b\to s\ell^+\ell^-\gamma$
can be calculated  directly using from the Hamiltonian of $b\to s\ell^+\ell^-$ in (\ref{hami}).
For example, contribution from the magnetic-penguin operator with {\it virtual} photon shown in
Fig. \ref{fig1} (a) reads:
\begin{eqnarray}
{\cal A}_a
&=& -i Gee_d\frac{m_{b}}{q^{2}}C_{7}^{eff}\left(\bar{s}\left[\frac{2p_{s}
   \cdot\epsilon+\spur\epsilon\not \! k}{p_{s}\cdot k}\gamma^{\mu}\not\! q
   P_{R}\right]b\right)\left[\bar{\ell}\gamma_{\mu}\ell\right],
\label{amp7a}
\end{eqnarray}
where $p_{b,s}, k$  denotes the momentum of quarks and photon respectively, $\epsilon$
is the vector polarization of photon and $G=\alpha_{em} G_FV_{tb}V_{ts}^*/(\sqrt{2}\pi)$,
 $e_d=-1/3$ is the number of electrical charge of the external quarks.
In deriving above equation, we have used
motion equation for quarks and $q^{\mu}\bar{\ell}\gamma_{\mu}\ell=0$.
Using Eqs.~(\ref{bmeson}) and (\ref{amp7a}),
we write the amplitude of $B_s\to \ell^+\ell^-\gamma$ at meson level as:
\begin{eqnarray}
A_a &=&2iGee_d\frac{m_{b}m_{B_s}}{q^{2}}C_{7}^{eff}
\frac{1}{p_{B_s}\cdot k}\int_{0}^{1}\frac{\phi_{B_s}(x)}{x}dx
\left[k_{\mu}q\cdot\epsilon-\epsilon_\mu k\cdot q
-i\epsilon_{\mu\nu\alpha\beta}\epsilon^{\nu} k^{\alpha}q^{\beta}\right]
\left[\bar{\ell}\gamma^{\mu}\ell\right],
\end{eqnarray}
where  $x$, $y=1-x$ are the momentum fractions shared
by $s$, $b$ quark in $B_s$. By doing the similar calculation to
diagrams (b)-(d), the decay amplitude is then obtained as:
\begin{eqnarray}
A_{a+b+c+d}&=&iGee_d\frac{1}{p_{B_s}\cdot k}\biggl\{
\left[C_1 i \epsilon_{\alpha\beta\mu\nu}p^\alpha_{B_s}\varepsilon^\beta{k^\nu}
+C_2p_{B_s}^\nu(\varepsilon_\mu k_{\nu}- k_\mu\varepsilon_\nu)\right]
\bar{\ell}\gamma^\mu{\ell}\nonumber\\
&&+C_{10}\left[C_+i\epsilon_{\alpha\beta\mu\nu}p^\alpha_{B_s}\varepsilon^\beta{k^\nu}
+C_-p_{B_s}^\nu(\varepsilon_\mu k_{\nu}-k_\mu\varepsilon_\nu)\right]
\bar{\ell}\gamma^\mu\gamma_{5}\ell\biggl\}.
\label{ampp}
\end{eqnarray}
The form factors in Eq. (\ref{ampp}) are found to be:
\begin{eqnarray}
C_1 &=&C_+\left(C_9^{eff} -2\frac{m_{b}m_{B_{s}}}{q^{2}}C_{7}^{eff}\right), \nn\\
C_2 &=&C_{9}^{eff}C_{-}-2\frac{m_{b}m_{B_{s}}}{q^{2}}C_{7}^{eff}C_{+},
\label{formfa}
\end{eqnarray}
where
\begin{eqnarray}
C_{\pm}=\int_{0}^{1}\left(\frac{1}{x}\pm\frac{1}{y}\right)\phi_{B_s}(x)dx.
\end{eqnarray} The expression in (\ref{ampp}) can be compared with Ref. \cite{LU06}.

Now we will focus attention on calculating the diagrams (e) and (f)
in Fig.~\ref{fig1} which are always neglected in other works.
In these two diagrams, photon of the magnetic-penguin operator is {\it real},
thus its contribution to $B_s\to \ell^+\ell^-\gamma$ is different from that of
magnetic-penguin operator with virtual photon in diagram (a) and (c).
We get the amplitude:
\begin{eqnarray}
A_{e+f}=i2Gee_dC_{7}^{eff}\frac{m_{b}m_{B_s}}{q^{2}}\frac{1}{p_{B_s}\cdot q}\overline{C}_+
\left[k_{\mu}q\cdot\epsilon-\epsilon_\mu k\cdot q-i\epsilon_{\mu\nu\alpha\beta}\epsilon^{\nu} k^{\alpha}q^{\beta}\right]
\left[\bar{\ell}\gamma^{\mu}\ell\right],
\end{eqnarray}
with coefficients $\overline{C_+}$ obtained by a replacement:
\begin{eqnarray}
\overline{C}_+&=& C_+(x\to \bar{x}=x-z-i\epsilon;~y\to \bar{y}=y-z-i\epsilon)\nn\\
&=& N_B\int_{0}^{1}dx({\frac{1}{x-z-i\epsilon}}+{\frac{1}{1-x-z-i\epsilon}})
x^{2}(1-x)^{2}\exp\left[-\frac{m_{B_s}^2}{2\omega_{B_s}^2}x^{2}\right],
\end{eqnarray}
where $z=\frac{q^{2}}{2p_{B_s}\cdot q}$. Note that  pole
in $\overline{C_+}$ corresponds to  the pole of the quark propagator when
it is connected by the off-shell photon propagator. Thus the  $\overline{C_+}$ term may enhance
the decay rate of $B_s\to \ell^+\ell^-\gamma$  and its  analytic expression reads
\begin{eqnarray}
\overline{C}_+
&=&2N_{B_s}\pi i z^{2}(1-z)^{2}\exp\left[-\frac{m_{B_s}^2}{2\omega_{B_s}^2}z^{2}\right]\nn\\
&+&N_{B_s}\int_{0}^{1}dx({\frac{1}{x+z}}-{\frac{1}{1+x-z}})
x^{2}(1+x)^{2}\exp\left[-\frac{m_{B_s}^2}{2\omega_{B_s}^2}x^{2}\right]\nn\\
&-&N_{B_s}\int_{-1}^{1}\left(\frac{1}{\frac{1}{x}-z}+\frac{1}{1-\frac{1}{x}-z}\right)
\frac{dx}{x^{4}}(1-\frac{1}{x})^{2}
\exp\left[-\frac{m_{B_s}^2}{2\omega_{B_s}^2}\frac{1}{x^{2}}\right].
\end{eqnarray}

As the contribution from the  Fig.\ref{fig1} (g) and (h) with photon
attached to external lepton lines, considering the fact that (i)
being a pseudoscalar meson, $B_s$ meson can only decay through axial current,
so the magnetic penguin operator $O_7$ 's contribution vanishes; (ii)
the contribution from operators $O_9,\ O_{10}$ has the helicity suppression
factor $m_{\ell}/m_{B_s}$, so for light lepton electron and muon,
we can neglect their contribution safely.

The total matrix element for the decay $B_s\to\ell^+\ell^-\gamma$ is
obtained a sum of the $ A_{a+b+c+d}$ and $A_{e+f}$.  After
summing over the spins of leptons and polarization of the photon,
and then performing the phase space integration over one of the two Dalitz variables, we get the
differential decay width versus  the photon energy $E_\gamma$,
\begin{eqnarray}
\frac{d\Gamma}{dE_\gamma}&=&\frac{\alpha^3{G^2_F}}{108\pi^4}|V_{tb}V^*_{ts}|^2
(m_{B_s}-2E_{\gamma})E_{\gamma}
\left[|\overline{C}_1|^2+|\overline{C_2}|^2+C_{10}^2(|C_+|^2+|C_-|^2)\right].
\end{eqnarray}
The coefficients $\overline{C}_i\ (i=1,2)$ can be obtained by a
shift:
\begin{eqnarray}
\overline{C}_i=C_i-\frac{2m_{b}m_{B_s}}{q^{2}}\frac{p_{B_s}\cdot k}{p_{B_s}\cdot q}\overline{C_{+}}C_7^{eff}.
\end{eqnarray}

\section{Results and discussions}\label{sec:num}
The decay branching ratios can be easily obtained by
integrating over photon energy. In numerical calculations,
we use the following parameters~\cite{PDG2012}:
 $$\alpha_{em}=\frac{1}{137},~ G_F = 1.166\times 10^{-5} {\rm GeV}^{-2},
 ~m_b=4.2{\rm GeV},$$
 $$|V_{tb}|=0.88,~|V_{ts}|=0.0387,~|V_{td}|=0.0084$$
 $$m_{B_s}=5.37{\rm GeV},~\omega_{B_s}=0.5,~f_{B_s}=0.24{\rm GeV},~\tau_{B_s}=1.47\times 10^{-12}s.$$
 $$m_{B_d^0}=5.28{\rm GeV},~\omega_{B_d} = 0.4,~f_{B_d} = 0.19{\rm GeV},~\tau_{B_d}= 1.53\times 10^{-12}s. $$
The ratios of $B_s\to \ell^+\ell^-\gamma$
with and without the contribution from  magnetic-penguin operator
with real photon are shown in Table \ref{table} together with results of
$B_{d,s}\to\ell^+\ell^-\gamma$  from this work and other models for comparison.
The errors shown in the Table~\ref{table} comes from the heavy meson wave function, by
varying the parameter $\omega_{B_d}=0.4\pm 0.1$, and
$\omega_{B_s}=0.5\pm 0.1$ \cite{LU06}.
Note that, the predicted branching ratios receive  errors from
many parameters, such as  meson decay constant, meson and quark masses etc.
\begin{table}[htbp]
\caption{Comparison of branching ratios with other model calculations}
\begin{center}
\begin{tabular}{c|c|c|c|c|c}
\hline\hline
Branching Ratios~($\times 10^{-9}$) &\multicolumn{2}{c|}{Our Results}&
\multicolumn{2}{c|}{Quark Model}& light cone \\
\hline
&Excluded Fig.1(e),(f)& Included Fig.1(e),(f)& Ref.\cite{LU06} &
 Ref.\cite{Eilam97}&Ref.\cite{Aliev97}\\
\hline
$B_s\to \ell^+\ell^-\gamma$ & $3.74_{-1.00}^{+1.76}$    & $7.45_{-1.82}^{+2.98}$  & 1.90   & 6.20  &  2.35 \\
$B_d^0\to \ell^+\ell^-\gamma$ & $0.16_{-0.05}^{+0.11}$ & $0.31_{-0.10}^{+0.20}$ & 0.08 & 0.82 & 0.15 \\
\hline\hline
\end{tabular}
\label{table}
\end{center}
\end{table}

A couple of remarks on the $B_s$ rare exclusive radiative decays are follows:
\begin{itemize}
\item[{\rm 1.}]
As pointed out in Ref.~\cite{li,LU06}, the branching ratios are
proportional to the heavy meson wave function squared, the radiative
leptonic decays are  very sensitive probes in extracting the heavy
meson wave functions;
\item[{\rm 2.}]
The contributions from magnetic-penguin operator  with real photon to
the exclusive decay is large, and  the branching of $B_s\to \ell^+\ell^-\gamma$
is  enhanced nearly by a factor 2 compared with that only contribution
from magnetic-penguin operator with virtual photon
and nearly up to $10^{-8}$, implying the search
of $B_s\to \ell^+\ell^-\gamma$  can be achieved in near future.
\item[{\rm 3.}]
Due to the large contributions from magnetic-penguin operator  with real photon, the
form factors for matrix elements $\langle\gamma|\bar{s}\gamma^\mu(1\pm\gamma_5) b|B_s\rangle$ and
$\langle\gamma|\bar{s}\sigma_{\mu\nu}(1\pm\gamma_5)q^\nu b|B_s\rangle$ as a function of dilepton mass
squared $q^2$ are not  as simple as $1/(q^2-q_0^2)^2$ where  $q_0^2$ is  constant \cite{Eilam95}.
The $B_s\to \gamma$  transition form factors predicted in this works have also some differences
from those in Ref. \cite{Melikhov04,Kruger03,Nikitin11}. For instance,
Ref. \cite{Kruger03} predicted the form factors $F_{TV}(q^2,0)$, $F_{TA}(q^2,0)$
induced by tensor and pseudotensor currents with emission of the virtual
photon, as shown in diagrams (a) and (c) of FIG. \ref{fig1},
are only equal at maximum photon energy, whereas the corresponding formula in this work have the same
expression  as $-\frac{e_dm_{B_s}}{p_{B_s}\cdot k} C_+\propto 1/(q^2-q_0^2)$ in Eq. (\ref{ampp}).
The research of $B_s\to \ell^+\ell^-\gamma$  may give some hints on these form factors.
\end{itemize}
At this stage, we  think it is necessary to  present a few more comments about the
calculation of Ref. \cite{Melikhov04}. In order to estimate the contribution of 
emission of the real photon from the magnetic-penguin operator,
the authors of Ref. \cite{Melikhov04} calculated the form factors
$F_{TA,TV} (0, q^2)$ by including the short distance contribution in $q^2\to 0$ limit
and additional long distance contribution from the resonances of vector mesons
such as  $\rho^0$, $\omega$ for $B_d$ decay and $\phi$  for $B_s$ decay.
Obviously, this means the pole mass enhancement of the valence quark
were  not appropriately taken into account. Moreover,
if $F_{TA,TV} (0, q^2)=F_{TA,TV} (0, 0)$ stands for the short distance
contribution, it seems double counting since in this case
photons which emits from magnetic-penguin vertex and quark lines directly
are not able to  be distinguished.

\section{Conclusion}\label{sec:con}
We evaluated  the rare decays
$B_s\to \gamma\ell^+\ell^-$ in the SM,  including two kinds of contributions from magnetic-penguin
 operator with virtual and real photon. In contrast to the previous works
which treated  contribution from  magnetic-penguins operators with real photon to the decays
 as negligible small, we found that the contributions is large, leading to  the branching of
 $B_s\to \ell^+\ell^-\gamma$ being  nearly enhanced by a factor 2.
In the current early phase of the LHC era, the exclusive modes with muons in the final states are
among the most promising decays. The decay $B_s\to \mu^+\mu^-$ is likely to be confirmed before
the end of 2012 \cite{Talk11}.  Although there are some theoretical challenges in calculation
of the hadronic form factors and non-factorable corrections, with the predicted branching
ratios at order of  $10^{-8}$,  $B_s\to\ell^+\ell^-\gamma$
can be expected as the next goal once $B_s\to \mu^+\mu^-$ measurement
is finished since the final states can be
identified easily and branching ratios are large. Our
predictions for such processes can be tested in the LHC-b and B factories in near future.
\begin{acknowledgments}
This work was supported in part by the NSFC No. 11005006, 11172008 and Doctor
Foundation of BJUT No. X0006015201102.
\end{acknowledgments}

\end{document}